\documentclass[aps,pre,twocolumn,amsmath,amssymb,10pt]{revtex4-1}

\usepackage{graphicx}
\usepackage{bm}

\begin{document}

\title{Collective Sedimentation of Squirmers under Gravity}

\author{Jan-Timm Kuhr}%
 \email{jan-timm.kuhr@tu-berlin.de}%
\author{Johannes Blaschke}%
\author{Felix R\"uhle}%
\author{Holger Stark}%
\affiliation{Institut f\"ur Theoretische Physik, Technische Universit\"at Berlin, Hardenbergstr.~36, 10623 Berlin, Germany}%

\date{\today}

\begin{abstract}
Active particles, which interact hydrodynamically, display a remarkable variety of emergent collective phenomena.
We use squirmers to model spherical microswimmers and explore the collective behavior 
of thousands of them under the influence of strong gravity using the method of multi-particle collision dynamics 
for simulating fluid flow.
The sedimentation profile depends on the ratio of swimming to sedimentation velocity as well as 
on the squirmer type. 
It shows close packed squirmer layers at the bottom and a highly dynamic region with exponential density 
dependence towards the top. 
The mean vertical orientation of the squirmers strongly depends on height. 
For swimming velocities larger than the sedimentation velocity, squirmers show strong convection in the exponential region. 
We quantify the strength of convection and the extent of convection cells by the vertical current density and its current dipole, 
which are large for neutral squirmers as well as for weak pushers and pullers.
\end{abstract}

\maketitle

\section{Introduction}
Microswimmers, whether biological or artificially produced, propel themselves forward without the help of any external force~\cite{Lauga:2009ku}.
Often however, external fields act on active particles~\cite{Stark:2016bm}.
Examples are microswimmers in shear~\cite{Sokolov:2009dy, Rafai:2010kg, Lopez:2015kk, Clement:2016kh}, 
Poiseuille~\cite{Rafai:2010kg, Zottl:2012el, Uppaluri:2012ja, Zottl:2013ij, Rusconi:2014et},
or swirling flow~\cite{Tarama:2014bv},
in harmonic traps~\cite{Nash:2010ir, Pototsky:2012ba, Hennes:2014hf, Menzel:2016kq},
and in light fields~\cite{Cohen:2014iy, Lozano:2016gd}.

The most natural example is gravity, which affects every swimmer that is not neutrally buoyant.
In dilute active suspensions, where hydrodynamic interactions between swimmers can be neglected, 
both experimental~\cite{Palacci:2010hk, Ginot:2015cv} 
and theoretical studies~\cite{Tailleur:2009ux, Enculescu:2011jz} find exponential density profiles similar to that of passive colloids, 
but with a sedimentation length $\delta$, which depends on activity and well surpasses that of passive colloids. 
Interestingly, in these dilute suspensions analytical studies show the emergence of polar order~\cite{Enculescu:2011jz},
and if the microswimmers are bottom heavy, the sedimentation profile can even be inverted~\cite{Wolff:2013dm}.

For higher densities of microswimmers hydrodynamic interactions become important and collective behavior 
emerges~\cite{Evans:2011wd, Marchetti:2013tq, Saintillan:2013wh, Alarcon:2013wv, Hennes:2014hf, Yan:2015fg, Zottl:2016vi, Stark:2016bm, Kruger:2016ju}.
This includes motility-induced phase 
separation~\cite{Cates:2015ft, Zottl:2014kh, MatasNavarro:2014fn, Navarro:2015ko, Blaschke:2016gf},
swarming~\cite{Cohen:2014iy, Oyama:2016eh}, and bioconvection~\cite{Wager:fp, Pedley:1992wa, Williams:2011cj, Williams:2011tt}, to 
name but a few phenomena. 
Furthermore, in real settings interactions with bounding surfaces are 
important~\cite{Llopis:2010wn, Schaar:2015kp, Zottl:2014kh, Lintuvuori:2016fl, Blaschke:2016gf}, especially if the swimmers are not perfectly 
buoyant~\cite{Ruhle:ztJgFmml}.

In this article we consider systems with thousands of microswimmers under the influence of gravity.
We simulate their full hydrodynamic flow fields using the method of multi-particle collision 
dynamics (MPCD)~\cite{Malevanets:1999ki, Gompper:2009ke}
in order to include hydrodynamic interactions between swimmers as well as between swimmers and bounding walls.
As a model microswimmer we use the squirmer~\cite{Lighthill:1952ta, Blake:1971ut, Ishikawa:2006vt, Downton:2009ig}, 
which is versatile enough to model the relevant swimmer types including pushers, pullers, and neutral swimmers.

In the following we concentrate on the case, where passive colloids would just strongly sediment to the bottom.
We show how density or sedimentation profiles depend on the ratio of active to sedimentation velocity as well as 
on the squirmer type. 
During collective sedimentation squirmers develop densely packed layers in the bottom region of the simulation cell. 
In contrast, we observe an exponential density profile in the upper region, where squirmers form a more dilute active suspension.
The mean vertical orientation of the squirmers depends strongly on their vertical position as well as on their swimmer type.
For swimming velocities larger than the sedimentation velocity, we find that hydrodynamic interactions organize
squirmers into convection cells. 
Importantly, both the strength of convection and the extension of the convection cells depend on the squirmer type.

The article is organized as follows.
We first introduce the squirmer as our model microswimmer and then shortly address the simulation method
of MPCD along with parameter settings and some details of our analysis in Section~\ref{sec:methodsandmodel}.
In Section~\ref{sec:results} we present our results of collectively sedimenting squirmers and analyze especially 
sedimentation and mean vertical orientation (in Section~\ref{sec:sedimentation}) 
and convection (in Section~\ref{sec:convection}).
Finally, in Section~\ref{sec:discussion} we summarize our findings and conclude.

\section{Methods and Model\label{sec:methodsandmodel}}

\subsection{The squirmer as a model swimmer}

In this work we use the squirmer~\cite{Lighthill:1952ta, Blake:1971ut, Ishikawa:2006vt, Downton:2009ig} as a model for a spherical microswimmer
with radius $R$.
It propels itself forward using a slip velocity field on its surface,
\begin{equation}
\label{eq:surface_field}
\bm{v}_s(\bm{r}_s) = B_1 \left( 1 + \beta \hat{\bm{e}} \cdot \hat{\bm{r}}_s \right) \left[ \left( \hat{\bm{e}} \cdot
\hat{\bm{r}}_s \right) \hat{\bm{r}}_s - \hat{\bm{e}} \right],
\end{equation}
which generates a hydrodynamic flow field in the surrounding fluid.
Here, $\bm{r}_s$ is a vector, which points from the center of the squirmer to a point on its surface, 
$\hat{\bm{r}}_s = \bm{r}_s / R$ is the corresponding unit vector, and the unit vector $\hat{\bm{e}}$ 
indicates the squirmer orientation.
In the bulk of a quiescent fluid the squirmer orientation coincides with its swimming direction.
Our model in eq.~\eqref{eq:surface_field} only takes into account the first two terms introduced in~\cite{Lighthill:1952ta, Blake:1971ut}.
They are sufficient to determine the swimming speed and swimmer type, by which microorganisms and 
artificial microswimmers like Janus particles~\cite{Golestanian:2005ka, Ebbens:2014be} or active droplets~\cite{Thutupalli:2011bv, Schmitt:2013jo, Schmitt:2016kv, Schmitt:2016vx, Maass:2016hz, Jin:iu}
are typically characterized.
Thus, the squirmer propels along its orientation vector $\hat{\bm{e}}$ with swimming speed
$v_0 = 2/3 B_1$ and creates fluid flow, the far field of which is controlled by the parameter $\beta$.
The value of $\beta$ therefore indicates the squirmer type.
While $\beta =0$ creates a neutral squirmer with the far field of a source dipole ($\sim r^{-3}$),
$\beta < 0$ and $>0$ refer to pushers or pullers, respectively, the flow fields of which decay like a force dipole
($\sim r^{-2}$)~\cite{Spagnolie:2012vn}.

\subsection{Multi-particle collision dynamics}

We investigate the behavior of many squirmers, which interact hydrodynamically with each other and with confining surfaces.
We employ multi-particle collision dynamics 
(MPCD)~\cite{Malevanets:1999ki, Padding:2006fw, Noguchi:2007fi, Kapral:2008dv, Gompper:2009ke}
to numerically solve the Navier-Stokes equations including thermal noise.
They reduce to the Stokes equations at low Reynolds numbers studied in this article.

In our MPCD simulations the fluid is modeled by ca.~$2\cdot10^7$ point particles of mass $m_0$.
Their positions $\bm{r}_i$ are updated in a streaming step using their velocities $\bm{v}_i$:
$\bm{r}_i(t + \Delta t) = \bm{r}_i(t) + \bm{v}_i\Delta t$. 
In the subsequent collision step fluid particles within cubic cells of linear extension $a_0$ exchange momentum according to 
the \mbox{MPC-AT+a} rule~\cite{Noguchi:2007fi}.
It conserves linear and angular momentum and also thermalizes velocities to temperature $T$.
Further details of our implementation are described in Refs.~\cite{Zottl:2014kh, Blaschke:2016gf}. 
We use here the parallelized version of Ref.~\cite{Blaschke:2016gf}, which is suited to simulate large systems with many swimmers.

In the present work, we consider squirmers under gravity. 
So we have to add an acceleration term $\bm{a} \Delta t^2 / 2$ to the squirmers position in the streaming step, where the 
acceleration $\bm{a}$ is due to the gravitational force $-mg \bm{e}_z$ along the vertical with $m$ the buoyant mass of a 
squirmer and $g$ the gravitational acceleration.
Since gravity does not induce a noticeable density change of the fluid on the micron length scale,
we do not apply a gravitational acceleration to the fluid particles. 
If a fluid particle encounters a bounding wall or a squirmer, the particle's position and velocity are updated according to the 
``bounce-back rule''~\cite{Padding:2005ez}, which implements either the no-slip boundary condition
or the surface flow field of eq.~\eqref{eq:surface_field}, respectively.
During the streaming step momentum is transferred from the fluid particles to the squirmers, the velocities of which are 
updated by a molecular dynamics step.
It includes steric interactions among squirmers and with bounding walls.

MPCD reliably reproduces  analytical results, 
including the flow field around passive colloids~\cite{Padding:2006fw},
the friction coefficient of a particle approaching a plane wall~\cite{Padding:2010fi},
the active velocity of squirmers~\cite{Downton:2009ig}, 
as well as the torque acting on them close to walls, where lubrication theory has to be applied~\cite{Schaar:2015kp}.
It also simulates correctly segregation and velocity oscillations in dense colloidal suspensions under 
Poiseuille flow~\cite{Kanehl:2015kw, Kanehl:2017fz}.

MPCD resolves flow fields on time and length scales large compared to the duration of the streaming step $\Delta t$ 
and the mean free path of the fluid particles, respectively.
Therefore, using a squirmer radius of $R = 4 a_0$, we expect to resolve hydrodynamic flow fields even when
squirmers are close to each other.

\subsection{Parameters}
We simulate the behavior of $N = 2560$ squirmers of radius $R = 4 a_0$ under gravity in a cuboidal box of extensions 
$L_x = L_y = 112a_0$ in the horizontal plane and $L_z = 224a_0$ along the vertical. 
Thus, the mean volume fraction of squirmers amounts to $\phi \approx 0.244$.
At $z = 0$ and $z = L_z$ our system is bounded by walls, while periodic boundary conditions apply in the horizontal directions.
For the duration of the streaming step we choose $\Delta t = 0.02 a_0\sqrt{m_0/k_B T}$, which sets the shear viscosity 
to $\eta = 16.05 \sqrt{m_0 k_B T}/a_0^2$~\cite{Noguchi:2008bm}.

In the following, an important parameter will be the ratio of active to bulk sedimentation velocity, 
\begin{equation}
\alpha := \frac{v_0}{v_g} \, .
\end{equation}
For spherical squirmers with buoyant mass $m$ and gravitational acceleration $g$ one has $v_g = mg / (6\pi \eta R)$.
We will keep $v_0$ fixed (it is set by $v_0=2/3 B_1$ with $B_1 = 0.1\sqrt{k_B T/m_0}$),
and choose three values of $mg$ in order to study the cases $\alpha = 0.3$, $1.0$, and $1.5$.
In addition, we mention the sedimentation lengths of passive Brownian particles with the same buoyant masses $m$, 
$\delta_0 = k_B T / (mg)$, where $k_B T$ is thermal energy. 
For the values of $\alpha$ given above we calculate from this formula 
the respective values $\delta_0 = 9.3\cdot10^{-4} R$, $3.1\cdot10^{-3} R$, and $4.7\cdot10^{-3} R$,
so that without activity squirmers would settle into a dense packing at the bottom of the simulation cell.

The P\'eclet number $\mathrm{Pe} = v_0 R / D$, where $D = k_BT / (6\pi \eta R)$ is the translational diffusion coefficient, 
has the value $\mathrm{Pe} = 323$ in all simulations, thus thermal translational motion is negligible. 
Furthermore, in all our simulations we have a Reynolds number of $\mathrm{Re} = v_0 R n_\text{fl}/\eta = 0.17$, 
where $n_\text{fl} = 10$ is the average number of fluid particles per collision cell.
This implies Stokesian hydrodynamics where inertia can be neglected. 
Finally, with the thermal rotational diffusion coefficient $D_r = k_BT / (8\pi \eta R^3)$,
we introduce the persistence number $\mathrm{Pe}_r = v_0/ (D_r R)$. 
It measures the distance in units of particle radius $R$, which the squirmer moves persistently in one direction, 
before rotational diffusion changes its orientation. 
For all our simulations we have $\mathrm{Pe}_r = 430$.
Thus, without gravity a single isolated squirmer would swim across the vertical extent $L_z$ of the simulation cell on
an almost straight line.

\subsection{Determining sedimentation lengths}

To determine the sedimentation lengths of the squirmers, we need to be sure that the system is in steady state.
In the beginning of the simulations we initialize the squirmers with random positions and orientations. 
We then observe that the collection of squirmers ``collapses'' towards the bottom wall by monitoring the mean squirmer 
height $\langle z \rangle$.
In continuing the simulations, we ensure that ultimately $\langle z \rangle$ does not show any deterministic trend,
but is only subject to fluctuations. We then simulate for a period of at least $10^4$ MPCD time units 
(i.e.~$10^4 a_0 \sqrt{m_0/k_B T}$) and use this simulation data for our further analysis.

As explained in the results section, we determine the sedimentation or density profile $\rho(z)$ of the squirmers, from which we
identify some layering at the bottom wall of the simulation box, which is followed by a transitional and then an exponential region. 
In the latter we determine the sedimentation length $\delta$ using an exponential fit. 
The difficulty is to specify a range of heights $[z_b,z_t]$, in which the exponential fit is performed. 
We have developed heuristic but robust criteria to identify this range. 
They ensure that neither layering at the bottom wall nor accumulation of squirmers at the top wall influences the fit values for $\delta$. 
As a first constraint we demand that $z_t$ is at least a distance of $10R$ away from the top wall.
For $z_b$ we require twice the height, which the squirmers would assume if they were all perfectly stacked in a hexagonal close packing.
Within this first specification for the range $[z_b,z_t]$ we then determine the final $z_t$ as the height, where the density $\rho$ is minimal.
For $z_b$  we take the smallest height $z$, where $\rho(z)$ falls below $8 \%$ of the hexagonal-close-packed density.
The value of $8 \%$ is a purely empirical value.

We use the data in the range $[z_b, z_t]$ to obtain the sedimentation length $\delta$ from exponential fitting.
In order to also estimate its error, we need to generate several estimates for $\delta$ from our data.
Therefore, we split up the simulation time after reaching steady state into ten intervals.
For each interval we perform exponential fits in four different ranges: (i) $[z_b, z_t]$, (ii) $[z_b+0.1\Delta z, z_t]$,
(iii) $[z_b, z_t - 0.1\Delta z]$, and (iv) $[z_b+0.1\Delta z, z_t-0.1\Delta z]$, where $\Delta z := z_t - z_b$.
We use the modified ranges as an additional measure to ensure that we are in the exponential regime.
As an estimate for $\delta$, we then take the mean of all 40 fits, while the corresponding standard deviation specifies the error.

\section{Results\label{sec:results}}
\begin{figure}
\centering
\includegraphics[width=0.85\linewidth]{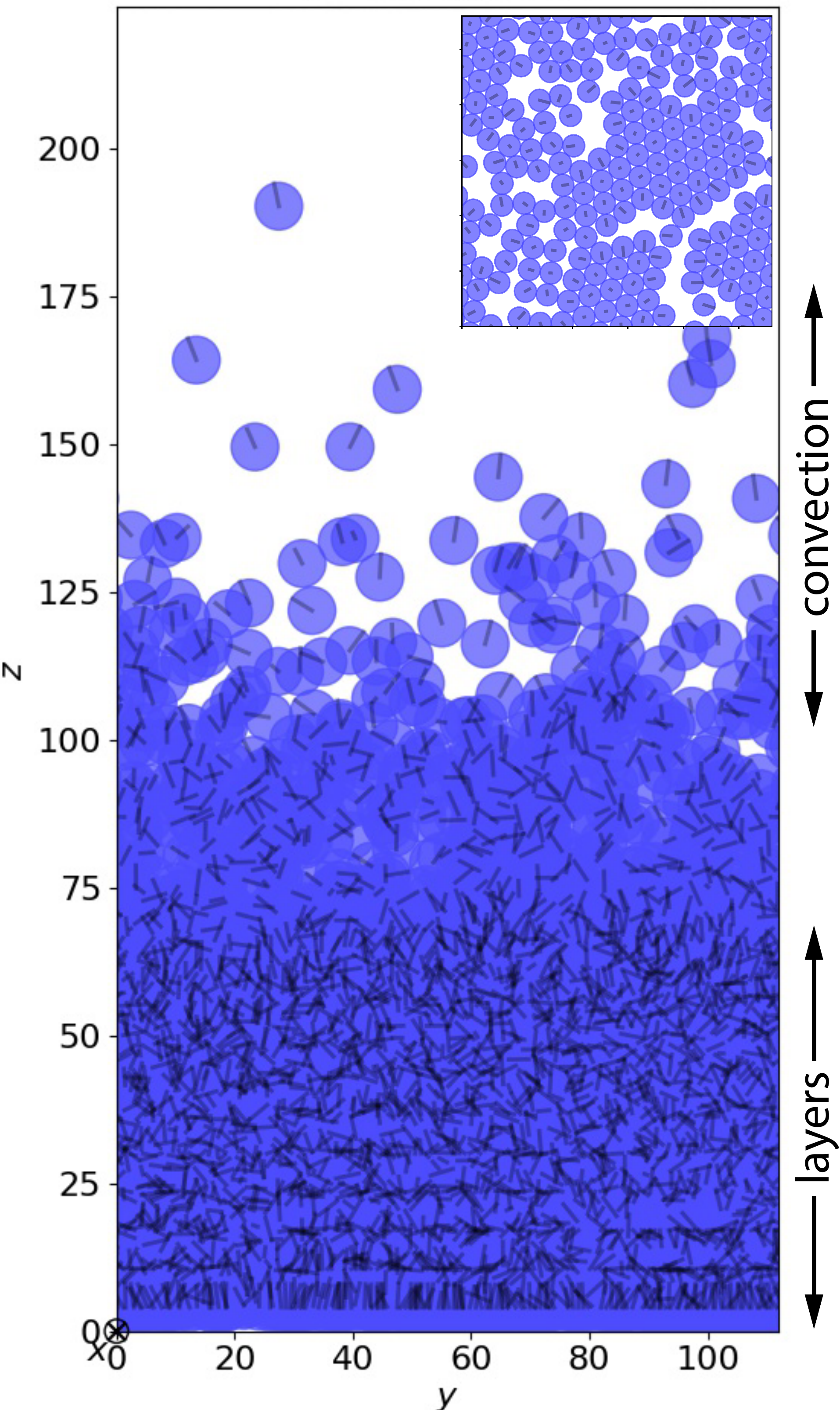}
\caption{%
Snapshot of 2560 neutral squirmers ($\beta =0$) moving under gravity in steady state at $\alpha = 1$.
The volume fraction is $\phi \approx 0.244$.
Three regions can be distinguished: layering at the bottom, followed by a transitional regime, and finally a region with
exponential density profile and convective motion of squirmers.
Inset: hexagonal clustering in the lowest layer of squirmers.
}
\label{fig:snapshot_sim_02}
\end{figure}

After a transient, the squirmers under gravity settle into a steady state, which we analyze in the following.
Figure~\ref{fig:snapshot_sim_02} shows a snapshot for a simulation, where the ratio of swimming to sedimentation velocity 
was $\alpha = 1.0$. In the lower part layers of squirmers have formed. 
In particular, the lowest layers display clusters of hexagonal packing (see inset of Fig.~\ref{fig:snapshot_sim_02}), 
which gradually dissolves when moving upward. 
After a transition region, where layering is not recognizable anymore, 
a dilute region of squirmers follows, where we will identify the exponential density profile.
In the supplemental material\dag, the two videos V1 and V2 (for $\alpha=1.5$) illustrate very impressively how dynamic the whole 
sedimentation profile is, especially in the exponential regime. 
This is in stark contrast to passive particles. 
In the following we will investigate some features of sedimenting squirmers in more detail.

\subsection{Sedimentation profile and vertical alignment\label{sec:sedimentation}}

\begin{figure}
\centering
\includegraphics[width=0.85\linewidth]{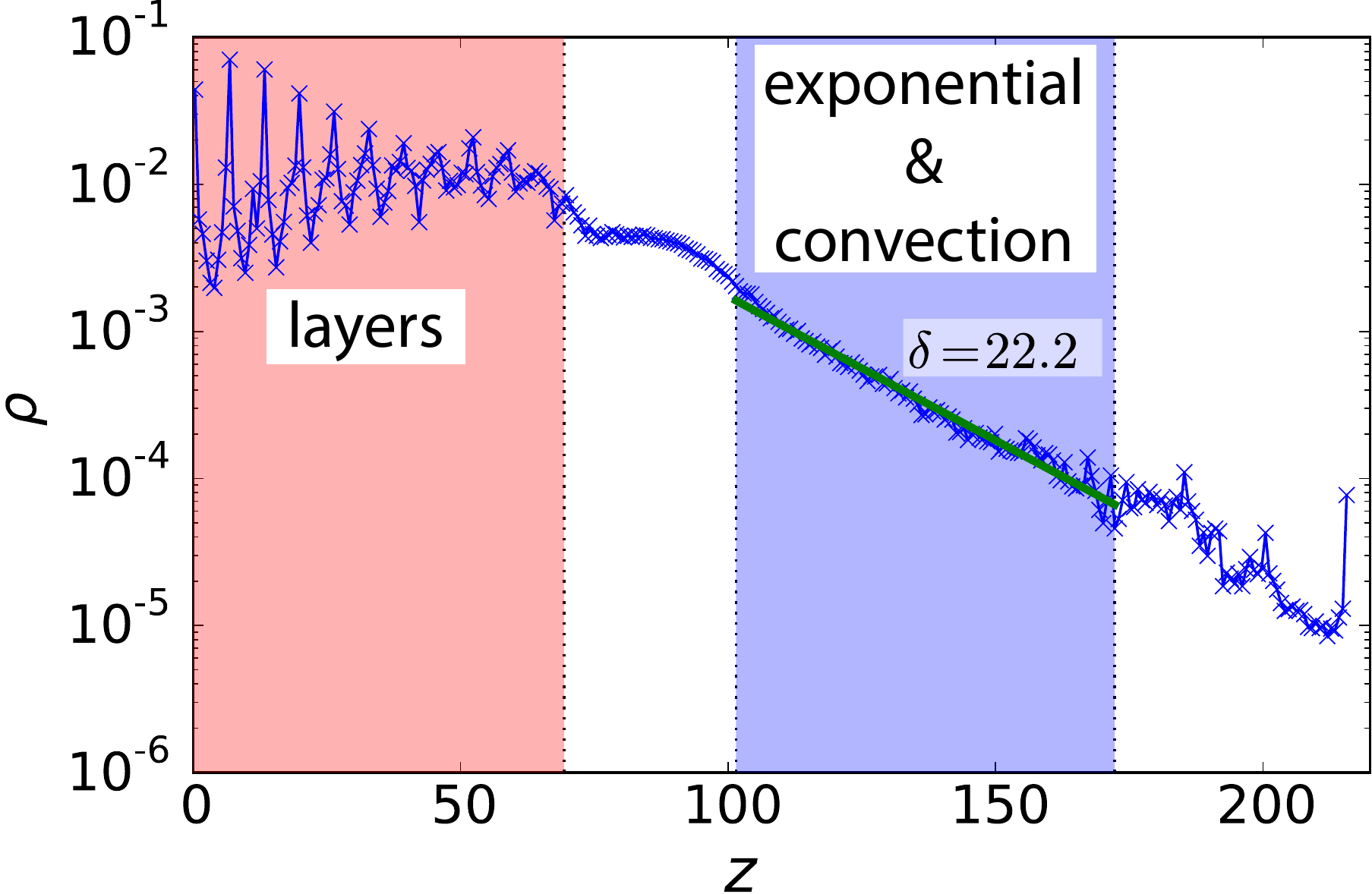}
\caption{%
Semi-logarithmic plot of the sedimentation profile $\rho(z)$ for the system 
in Fig.~\ref{fig:snapshot_sim_02} ($\alpha=1.0$ and $\beta=0$).
Different regions are indicated.
The green line is an exponential fit to extract the sedimentation length $\delta$.
}
\label{fig:hist02}
\end{figure}

\begin{figure*}
\centering
\includegraphics[width=\textwidth]{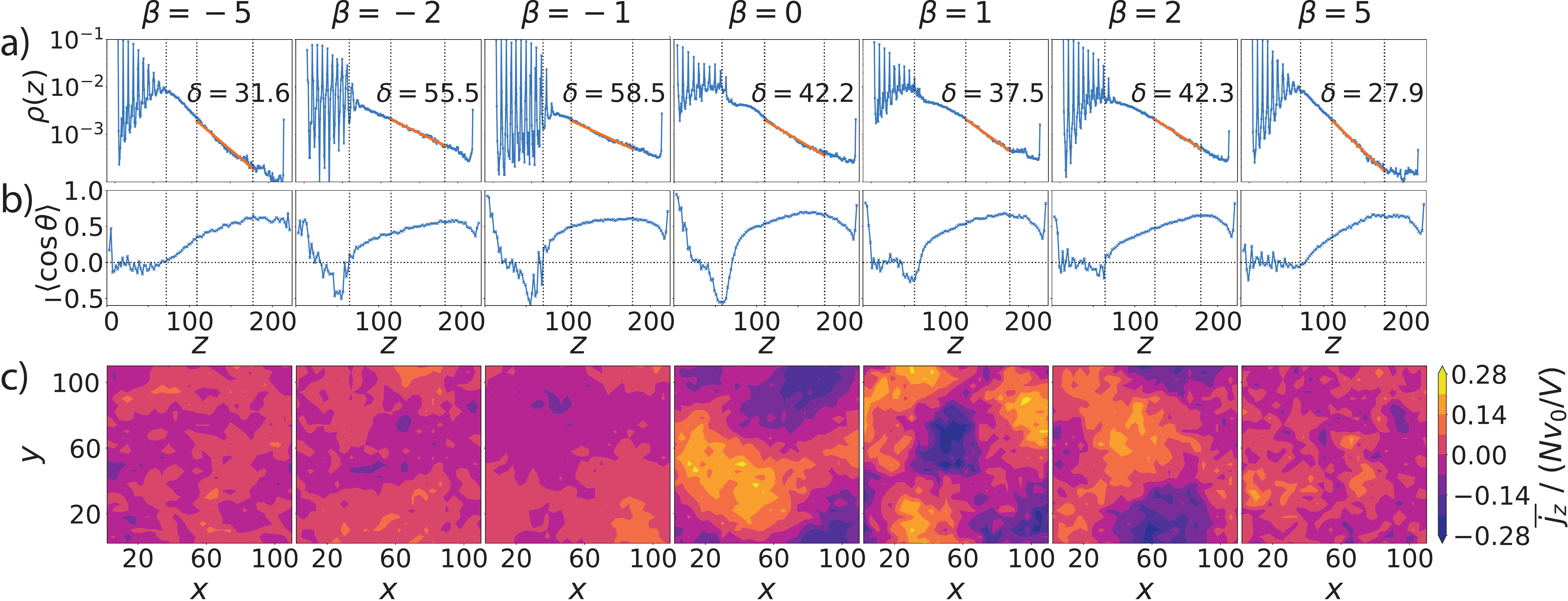}
\caption{%
Parameter study for varying $\beta$ at $\alpha=1.5$.
a)~Semi-logarithmic plot of the squirmer density $\rho(z)$ as a function of height $z$. 
Vertical dashed lines in each plot indicate (from left to right) top of layering, followed by the start and end of the exponential regime, 
where the sedimentation length $\delta$ is extracted by an exponential fit (bold orange line).
b)~Mean vertical orientation of squirmers as a function of height $z$.
$\theta$ denotes the angle between the vertical and the squirmer orientation.
c)~Vertical squirmer current density $\overline{j_z(\bm{x})}$, 
averaged over the exponential regime and time, color-coded in the $xy$ plane.
}
\label{fig:parameter_study}
\end{figure*}

The sedimentation profile in Fig.~\ref{fig:hist02} quantifies the observations from Fig.~\ref{fig:snapshot_sim_02}.
Close to the bottom, layering is clearly observable in the volume fraction or density $\rho(z)$ and indicated by peaks, the height of
which gradually decreases within ca.\ 11 layers. 
After a transitional regime, $\rho(z)$ decays exponentially and then is influenced by the upper bounding wall. 
Passive Brownian particles with buoyant mass $m$ show an exponential sedimentation profile 
with sedimentation length $\delta_0 = k_BT / mg$, as previously introduced.
For very dilute suspensions of active particles, one can still 
derive~\cite{Tailleur:2008kd, Tailleur:2009ux, Palacci:2010hk, Enculescu:2011jz, Wolff:2013dm, Stark:2016bm} 
and observe~\cite{Palacci:2010hk, Ginot:2015cv} an exponential profile, however, with an increased sedimentation length 
$\delta > \delta_0$. 
Even if passive particles all sink to the bottom due to their weight ($\delta_0 \ll R$), 
active particles with sufficiently large swimming speed can rise from the bottom with a sedimentation length $\delta > R=4$. 
In our simulations, squirmers strongly interact hydrodynamically by the flow fields they generate. 
Nevertheless, we observe an exponential decay of the density $\rho(z)$ similar to lattice-Boltzmann 
simulations~\cite{Cates:kh, Nash:2010ir} and experiments~\cite{Palacci:2010hk, Ginot:2015cv}, which we find non-trivial. 
Thus from fits to the exponential part of the density profile, $\rho(z) \sim e^{-z/\delta}$, we extract the sedimentation length $\delta$.

\begin{figure}
\centering
\includegraphics[width=\linewidth]{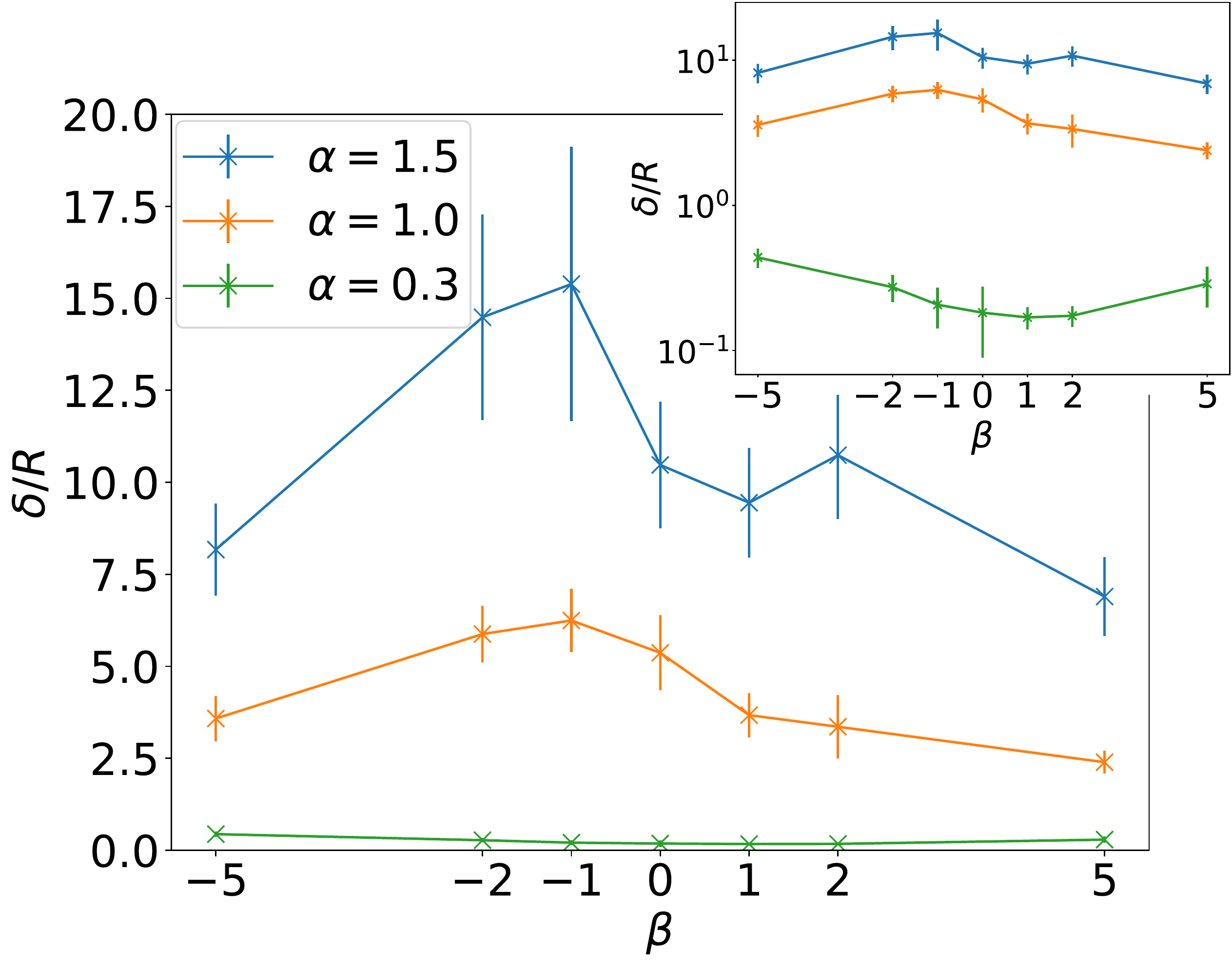}
\caption{%
Sedimentation length $\delta$ in units of $R$ as a function of squirmer parameter $\beta$ for different ratios of swimming to sedimentation 
velocities, $\alpha$.
Inset: semi-logarithmic plot.}
\label{fig:seg_len_vs_beta}
\end{figure}
In Fig.~\ref{fig:parameter_study}a) we present sedimentation profiles for a larger swimming speed, $\alpha = 1.5$, and different 
swimmer types $\beta$ to explore the influence of pushers and pullers.
At larger $\alpha$, all profiles show an exponential regime with larger sedimentation length compared to $\alpha =1$.
Fig.~\ref{fig:seg_len_vs_beta} shows a parametric study of $\delta$ in units of squirmer radius $R$ plotted versus $\beta$ and for three values of $\alpha$. 
Clearly, $\delta$ decreases with $\alpha$ and is only a fraction of $R$ for $\alpha = 0.3$, when activity is too small for particles to swim upwards. 
However, already the ratio $\alpha = 1.5$ is sufficient to have sedimentation lengths $\delta \approx 10 R$.
To address the robustness of our results, for the case $\beta = 0$ and  $\alpha = 1.5$, we reduced the volume density $\phi$ 
by decreasing the number of squirmers $N$ to a value where layer formation does not occur,
while keeping the height $L_z$ of the simulation box fixed.
This does not influence the sedimentation length $\delta$ significantly, 
as long as the squirmer density at the bottom of the box is similar to that of the top layers.
For $\beta = 0$ we also reduced both $N$ and $L_z$ by a factor of two, which keeps $\phi$ constant.
We find that the sedimentation length is reduced by about 30\% and 40\% for $\alpha = 1$ and $\alpha = 1.5$, respectively.
This is not surprising, since hydrodynamic interactions with the top wall, which were not relevant before, 
push squirmers downwards and also turn them away from the wall~\cite{Spagnolie:2012vn, Berke:2008cg}, which 
makes them swim downwards. 

In Fig.~\ref{fig:seg_len_vs_beta} we realize that for 
weak pushers ($\beta = -1$) the sedimentation length is largest and decreases for stronger pushers and also pullers. 
It has been reported in literature that the interaction between parallel squirmers grows with $|\beta|$~\cite{Ishikawa:2006vt, Gotze:2010cs}.
We speculate that the collective interactions of many squirmers will therefore stronger randomize swimming directions and hinder
squirmers with large $|\beta|$ from reaching larger heights, as reflected in the sedimentation length.
The larger sedimentation length for weak pushers as compared to neutral squirmers is, however, unexpected.
A possible explanation comes from the shape of flow fields for $\beta \neq 0$.
Pushers, in their center-of-mass frame, have a stagnation point  with vortical flow in front of them, 
while pullers have it at their back~\cite{Evans:2011wd}.
Since squirmers are typically pointing up in the exponential regime (see below) and since their density $\rho(z)$ decreases with height $z$,
pullers reorient more nearby squirmers 
compared to pushers, which decreases $\delta$.
Interestingly, the trend of $\delta$ for varying $\beta$ is inverted for small $\alpha$ with the minimum being at $\beta \approx 1$.
The reason for the inversion is not clear, but since $\delta < R$, we assume that interactions with the densely packed squirmer layers are relevant. 

In Fig.~\ref{fig:parameter_study}a) we also observe how the layer structure in the lower part of the system is influenced by $\beta$.
For $\beta = 0$ and $\beta = 1$ the minima between successive layers are less pronounced, which implies less order.
In contrast, especially for $\beta = -1$ and $\beta = -2$ layering is more pronounced.
We ascribe this difference to the hydrodynamic interactions between neighboring squirmers, which depend on $\beta$.

\begin{figure}
\centering
\includegraphics[width=.9\linewidth]{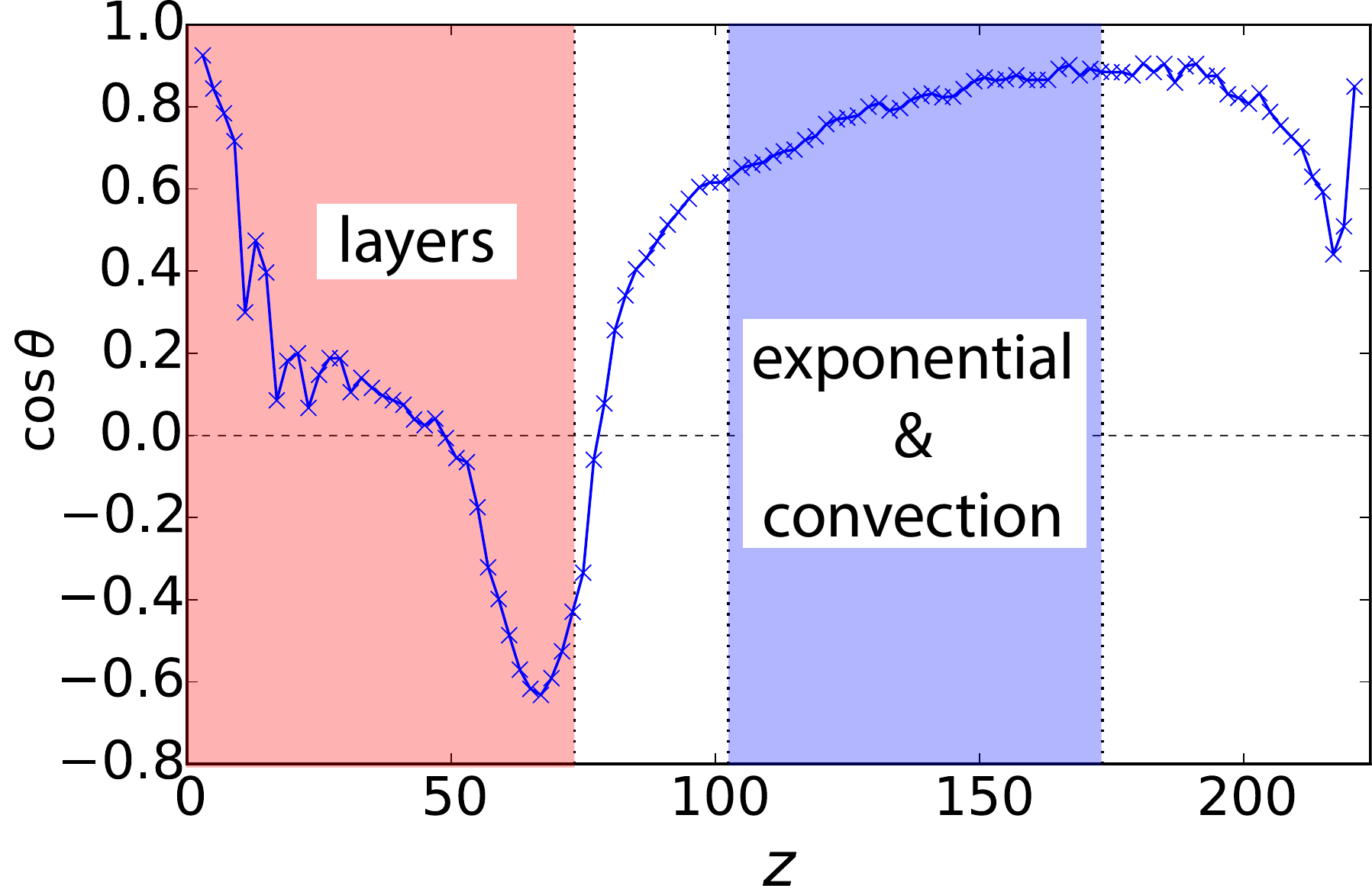}
\caption{%
Mean vertical orientation of squirmers, $\langle \cos\theta \rangle$, as a function of height $z$ for the system in Fig.~\ref{fig:snapshot_sim_02}
($\alpha=1.0$ and $\beta=0$).
}
\label{fig:orientationvsz}
\end{figure}

Finally, in Fig.~\ref{fig:orientationvsz} we show the mean orientation of squirmers as a function of height $z$ for the system
illustrated in Fig.~\ref{fig:snapshot_sim_02}.
The neutral squirmers in the bottom layer at $z=0$ have an upright orientation due to hydrodynamic interactions with the 
bounding wall~\cite{Ishikawa:2006vt, Zottl:2014kh, Schaar:2015kp, Lintuvuori:2016fl}.
The mean orientation then decreases to zero (see also Fig.~\ref{fig:snapshot_sim_02}) and drops to a negative value at the 
rim of the layering. This is simply because squirmers from above swim into the dense squirmer region and need some time to 
reorient and swim away. In the transitional region the orientation changes again rapidly to nearly upright and shows only small 
variations in the exponential regime.

The occurrence of polar order in the sedimentation profile of dilute suspensions has been predicted by 
theory and already occurs without any interactions
(i.e.~for dilute suspensions of active Brownian particles) 
just for kinetic reasons~\cite{Enculescu:2011jz,Stark:2016bm}. 
Hydrodynamic interactions between squirmers obviously do not destroy the polar order. 
In our parametric study of Fig.~\ref{fig:parameter_study}c), this is also confirmed for other squirmer types $\beta$. 
Differences occur in the layering and in the transitional region. 
For pullers the upright orientation in the bottom layer decreases for larger $\beta$, as expected by hydrodynamic 
interactions with the bottom wall in lubrication theory~\cite{Ishikawa:2006vt, Schaar:2015kp, Lintuvuori:2016fl}. 
In the adjacent layers hardly any polar order is visible in contrast to neutral squirmers. 
Weak pushers ($\beta = -1$ and $- 2$) show a similar but weaker trend compared to neutral squirmers. 
For strong pushers ($\beta=-5$), again, there is hardly any polar order in the layering.

\subsection{Convection\label{sec:convection}}
\begin{figure}
\centering
\includegraphics[width=0.9\linewidth]{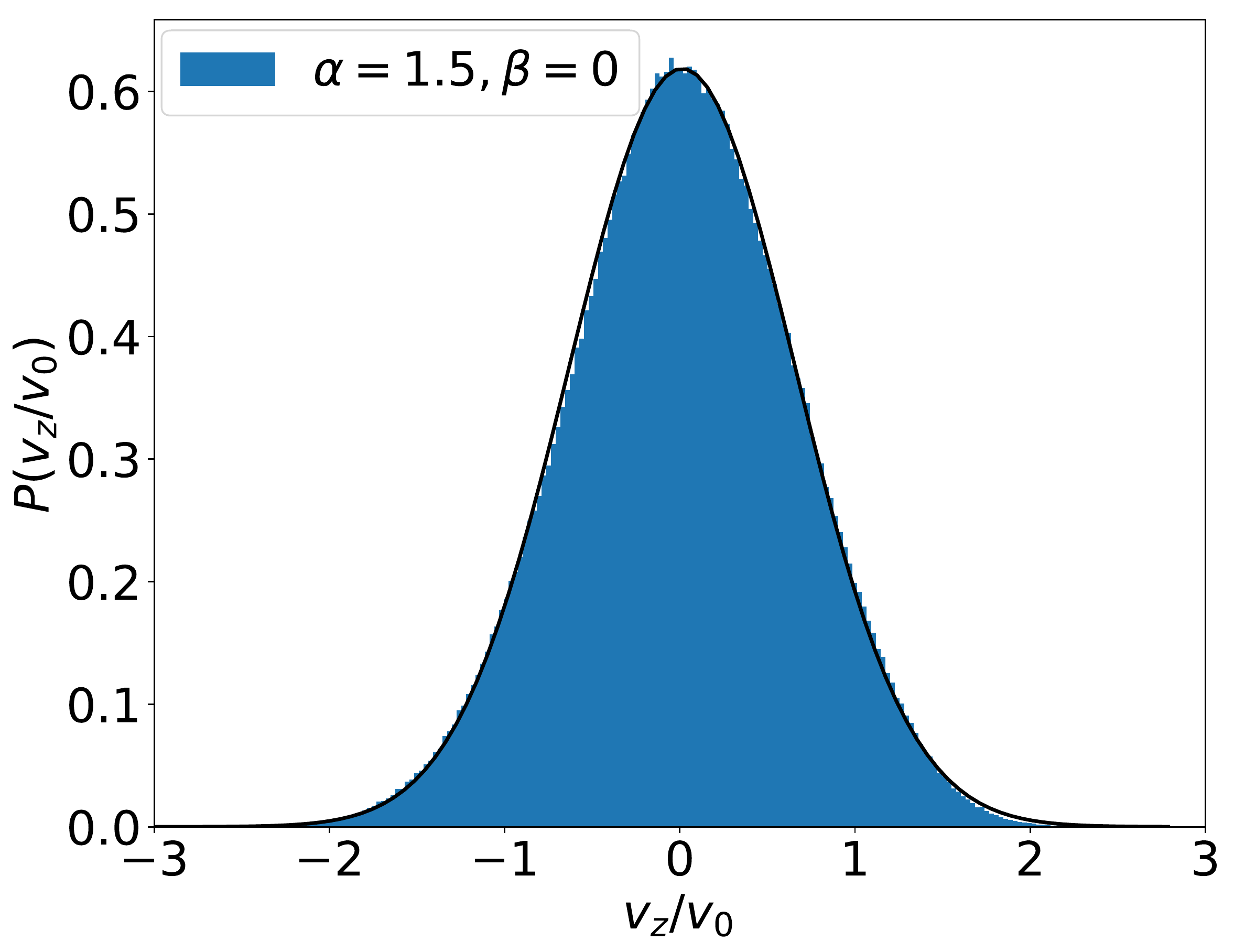}
\caption{%
Distribution of vertical squirmer velocity (blue) and Gaussian fit (black)
for $\alpha=1.5$ and $\beta = 0$ in the exponential regime.}
\label{fig:hist_vertical_velocity_exp}
\end{figure}

As videos V1 and V2 in the supplemental material\dag{} demonstrate, the squirmers in the exponential density region are very
mobile. 
In fact, while their mean vertical velocity is zero, the steady-state distribution of vertical velocities for $\beta = 0$ and 
$\alpha = 1.5$ can be well fitted by a Gaussian with standard deviation comparable to $v_0$ 
(see Fig.~\ref{fig:hist_vertical_velocity_exp}).
For large $|\beta|$ small deviations from the Gaussian form occur.
Thus, also vertical squirmer speeds larger than 
$\frac13 v_0$ arise (the maximal vertical velocity a single bulk squirmer can have at $\alpha = 1.5$). 
This means that squirmers are advected by flow fields set up by their neighbors. 
Indeed, for neutral squirmers and weak pullers/pushers, we see evidence for convection flow extending over the whole simulation cell.

\begin{figure}
\centering
\includegraphics[width=0.9\linewidth]{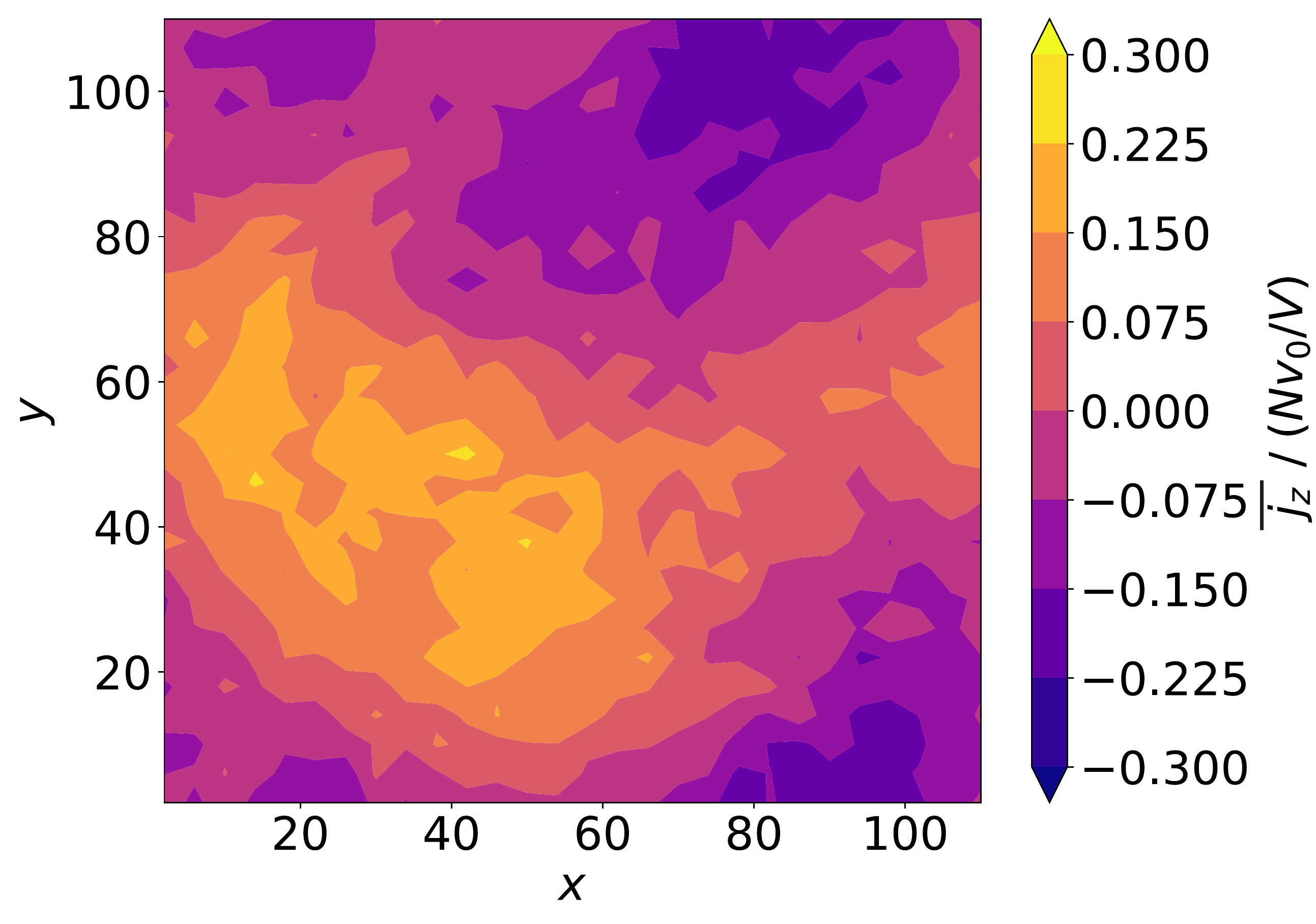}
\caption{%
Squirmer current density, $\overline{j_z(\bm{x})}$, averaged over the exponential regime and time, 
color-coded in the $xy$ plane for $\alpha=1.5$ and $\beta = 0$.
}
\label{fig:spatial_j}
\end{figure}

To quantify convection, we take the vertical squirmer current density and average it along the vertical in the 
exponential regime:
\begin{equation}
j_z(\bm{x}) = \langle \rho \rangle_{\|}(\bm{x}) \langle v_z \rangle_{\|}(\bm{x})
\end{equation}
where $\bm{x}=(x,y)$ is a position in the horizontal plane, $\langle \ldots \rangle_{\|}$ means average along the 
vertical, and $\rho$ is the squirmer density.
In the following we always indicate a time average over some quantity $q$ in the steady state by $\overline{q}$.
We plot $\overline{j_z(\bm{x})}$ in Fig.~\ref{fig:spatial_j} for neutral squirmers and $\alpha=1.5$ in the $xy$ plane of the
simulation box. 
While in the lower left squirmers move upward, the vertical current goes downward in the upper right indicating a convection 
cell, which extends over the whole horizontal plane. 
In Fig.~\ref{fig:parameter_study}c) we present $\overline{j_z(\bm{x})}$ for different squirmer types at the same $\alpha$. 
For weak pullers ($\beta=1$ and $2$) the extent of the convection cell decreases 
and at $\beta = 5$ large-scale convection is no longer observable.
For weak pushers ($\beta=-1$ and $-2$) the current density becomes weaker but still large-scale convection is visible, 
which then vanishes for $\beta=-5$.
For the case $\beta = 0$, $\alpha = 1.5$ we checked that large-scale convection is stable against a reduction in 
the squirmer density $\phi$ at constant $L_z$ and for reduced $L_z$ while keeping $\phi$ constant.
In both settings a single convection cell extends across the simulation box.
\begin{figure}
\centering
\includegraphics[width=0.9\linewidth]{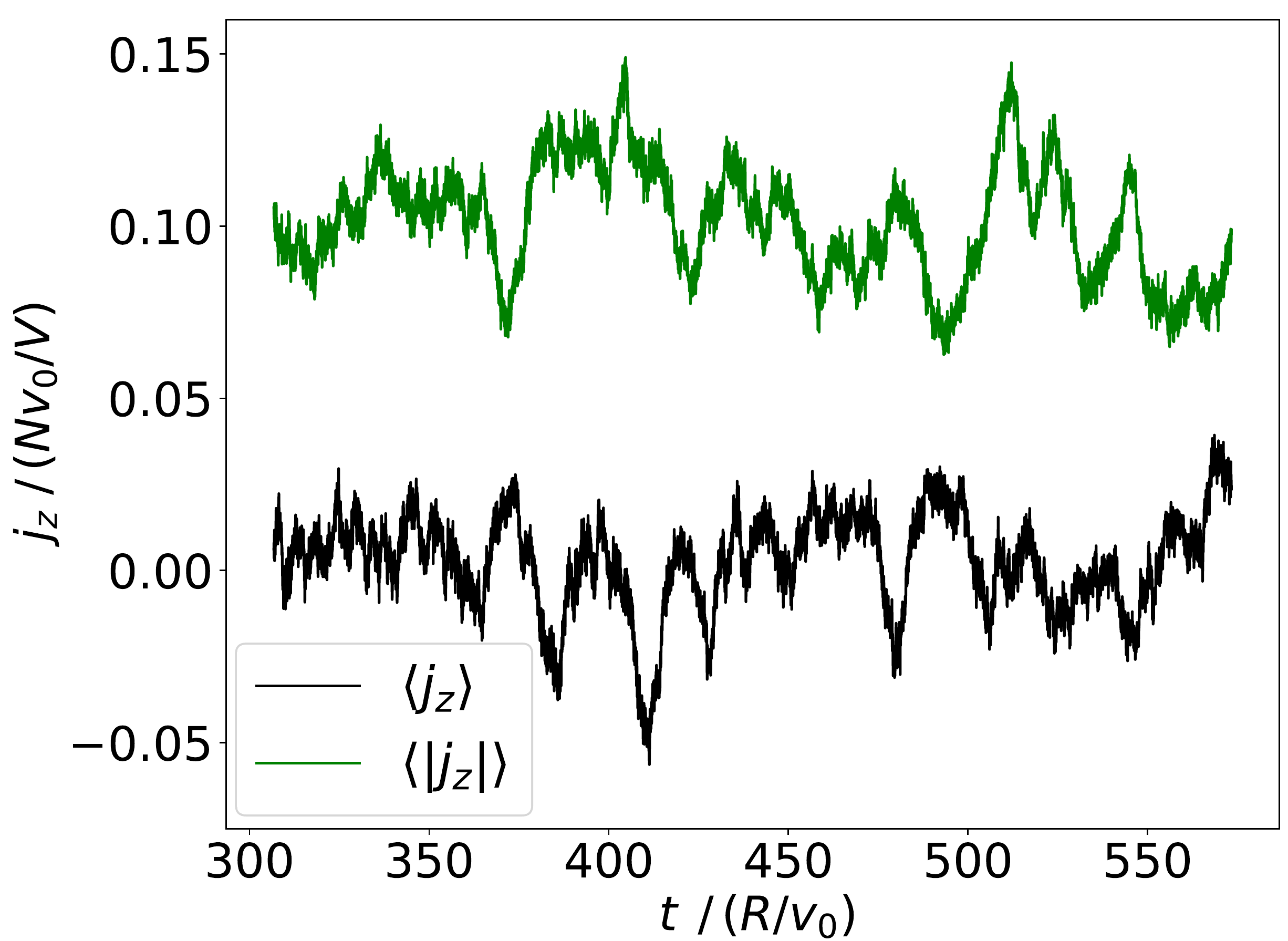}
\caption{%
Volume average of the squirmer current density, $\langle j_z \rangle$, and its magnitude, $\langle | j_z | \rangle$,
plotted versus time for $\alpha = 1.5$ and $\beta = 0$.
}
\label{fig:currents_vs_t}
\end{figure}
\begin{figure}
\centering
\includegraphics[width=0.9\linewidth]{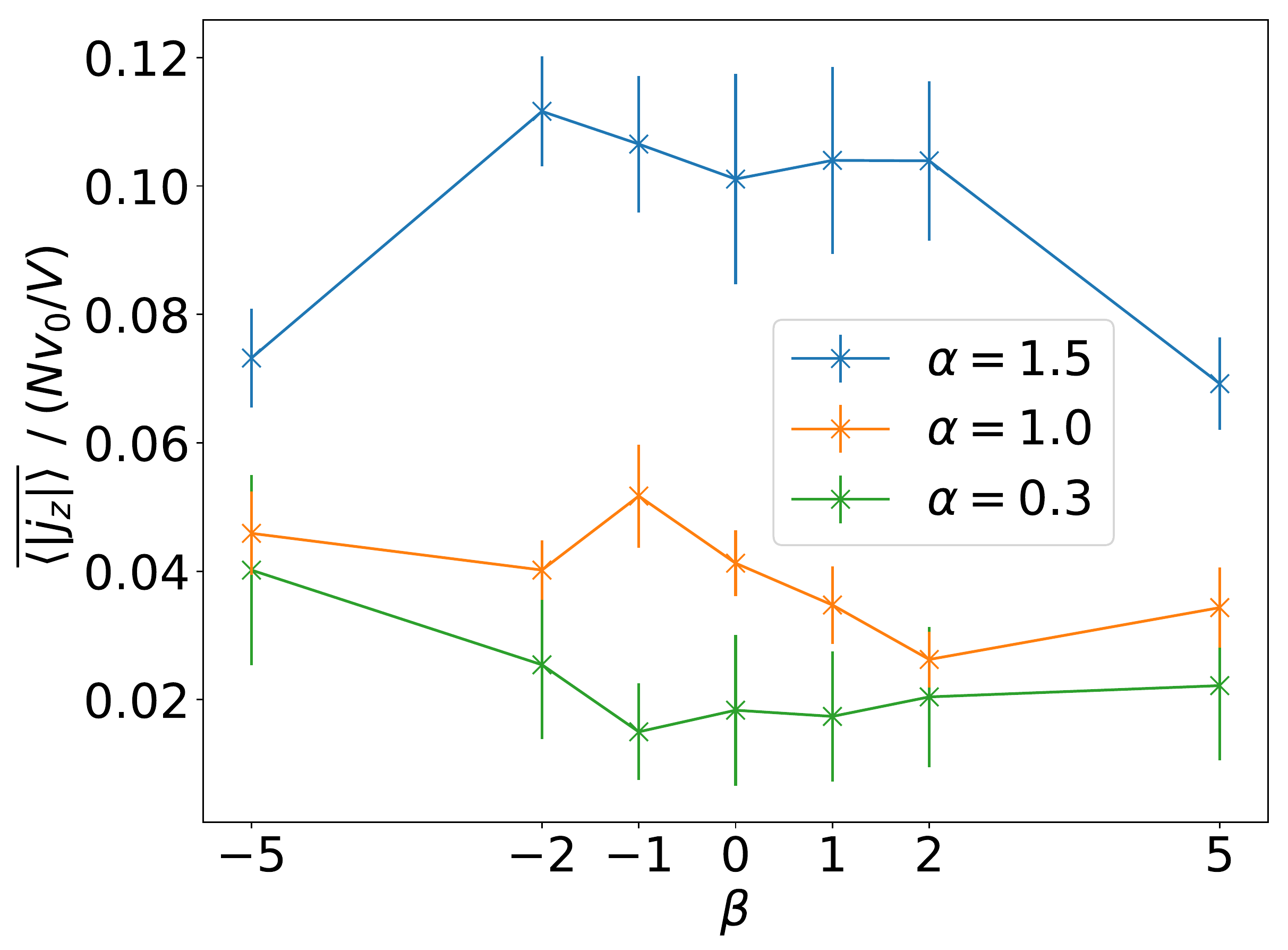}
\caption{%
Time and volume average of the magnitude of the current density, $\overline{\langle | j_z | \rangle}$,
as a function of squirmer parameter $\beta$ for different rescaled swimming speeds $\alpha$.
}
\label{fig:jsquared_vs_beta}
\end{figure}

To further characterize the squirmer density current $j_z(\bm{x})$, we calculate its zeroth and first moment.
The zeroth moment is the volume average of the vertical current density:
\begin{equation}
\langle j_z \rangle = \frac{1}{A} \int_A j_z(\bm{x}) d^2x = \langle \rho \rangle \langle v_z \rangle \, ,
\end{equation}
where $A$ is the area of the $xy$ plane of the simulation cell and $\langle \ldots \rangle$ means average over the whole
volume of the exponential region. 
The vertical current density $\langle j_z \rangle$ strongly fluctuates in time (see Fig.~\ref{fig:currents_vs_t}) but in steady 
state its temporal mean, $\overline{\langle j_z \rangle}$, has to be zero because of particle conservation. 
Thus, we use $\langle | j_z | \rangle$ to quantify how mobile the squirmers are in the exponential region 
(see Fig.~\ref{fig:currents_vs_t}). 
In Fig.~\ref{fig:jsquared_vs_beta} we show the time average $\overline{\langle | j_z | \rangle}$ versus $\beta$ for different 
rescaled swimming velocities $\alpha$. 
As expected, at $\alpha > 1$ the squirmers are more mobile than for $\alpha \le 1$ since they are able to move against gravity.
Furthermore, for $\alpha = 1.5$ the mean vertical current density $\overline{\langle | j_z | \rangle}$ decreases for large $| \beta |$,
revealing again the importance of the advective flow set up by the different squirmer types.

We call the first moment of $j_z(\bm{x})$ current dipole,
\begin{equation}
\bm{j}_D = \frac{1}{A} \int_A \bm{x} j_z(\bm{x}) d^2x \, ,
\label{eq:def_dipole_dens}
\end{equation}
where we choose the center of the $xy$ plane as the origin of $\bm{x}$.

\begin{figure}
\centering
\includegraphics[width=0.9\linewidth]{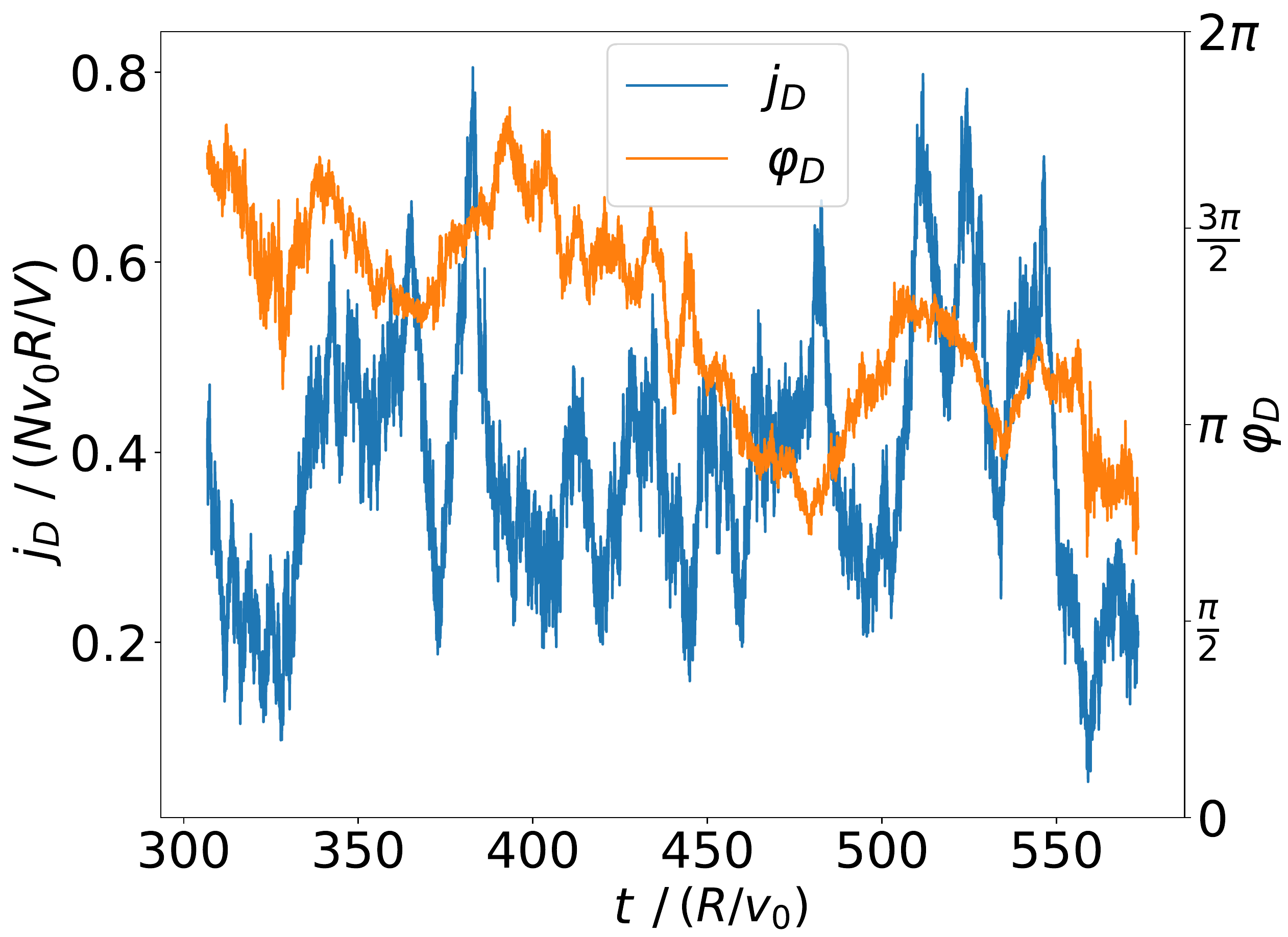}
\caption{%
Magnitude $j_D$ and orientation angle $\varphi_D$ of the current dipole plotted versus time for $\alpha=1.5$ and $\beta = 0$.
}
\label{fig:dipole_magnitude_and_angle_vs_t_03}
\end{figure}

The current dipole serves to quantify the strength and horizontal extension of the convection cell by its magnitude 
$j_D := |\bm{j}_D|$. 
The cell's orientation relative to the $x$-axis is given by the angle $\varphi_D$ with  $\cos \varphi_D = \bm{j}_D \cdot \bm{e}_x / j_D$. 
Figure~\ref{fig:dipole_magnitude_and_angle_vs_t_03} shows how $j_D$ strongly fluctuates in time, reflecting again the high mobility of the 
squirmers. 
The orientation angle $\varphi_D$ also fluctuates and in the example of Fig.~\ref{fig:dipole_magnitude_and_angle_vs_t_03} assumes two 
mean orientations around $1.5 \pi$ and $\pi$.
Overall, we can record that the spatial arrangement of convection is subject to strong fluctuations and strongly variable in time. 

\begin{figure}
\centering
\includegraphics[width=0.9\linewidth]{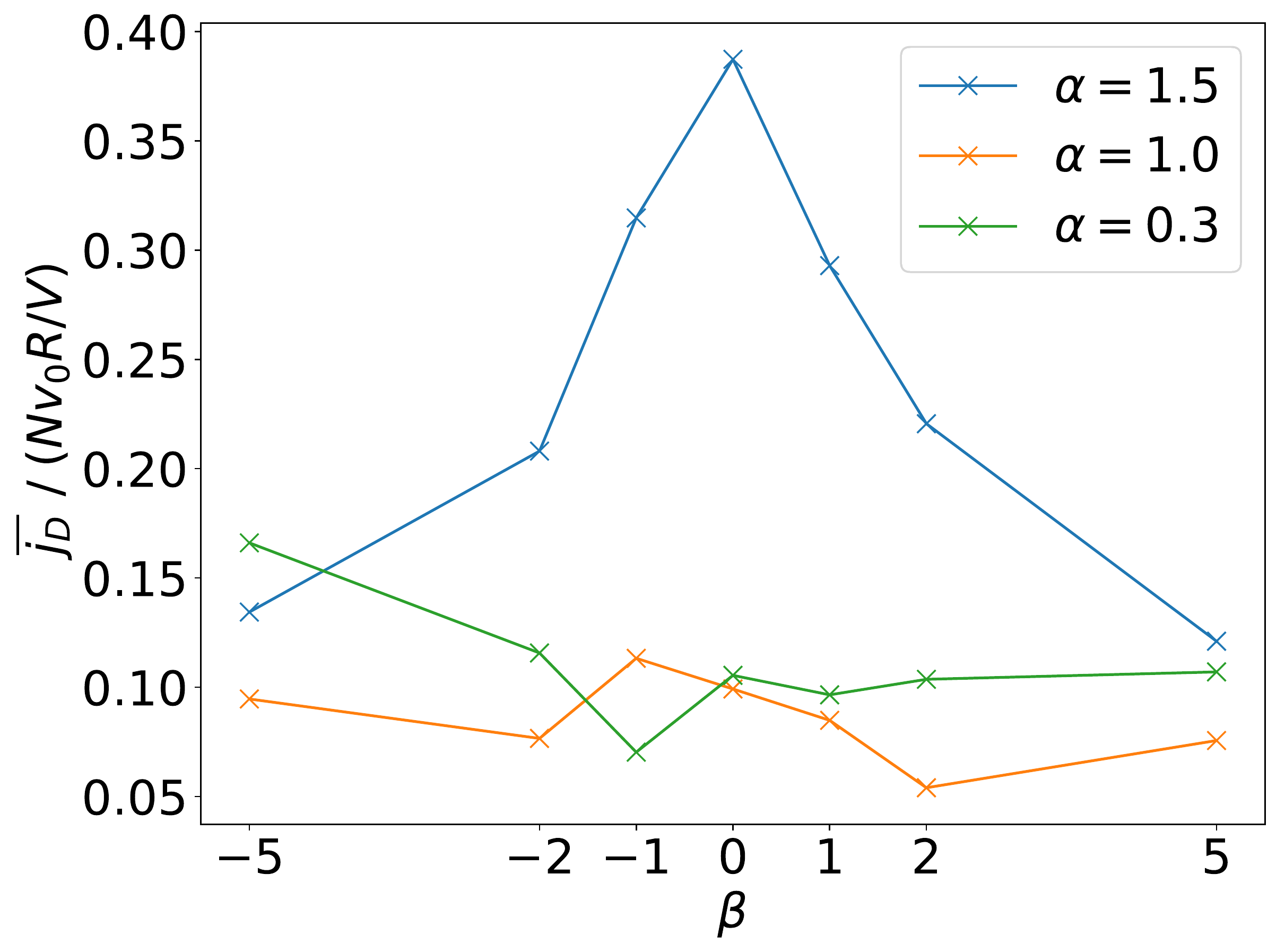}
\caption{%
Time average of the magnitude of the current dipole, $\overline{j_D}$,
as a function of squirmer parameter $\beta$ for different rescaled swimming speeds $\alpha$.
}
\label{fig:dipole_dens_vs_beta}
\end{figure}

Finally, in Fig.~\ref{fig:dipole_dens_vs_beta} we plot the time average $\overline{j_D}$ versus squirmer type $\beta$
for different $\alpha$. 
Convection is largest for large $\alpha$ and neutral squirmers. 
The current dipole we define in Eq.~\eqref{eq:def_dipole_dens} can be interpreted in analogy with a charge dipole in electrostatics.
Its magnitude changes when either the current density (the separated ``charges'') changes in magnitude or when
the distance between regions of positive vs. negative vertical speed is altered (the distance of ``charge separation'').
In accordance with Fig.~\ref{fig:parameter_study}c) we find that $\overline{j_D}$ for $\alpha=1.5$ decreases when the 
magnitude of the current density $j_z(\bm{x})$ decreases (weak pusher, $\beta <0$) or the extension of the the convection 
cell becomes smaller (weak puller, $\beta>0$).
Strong pushers/pullers with $|\beta| = 5$ only show weak convection.
For small rescaled swimming speed $\alpha \le 1$ convection is generally small.

\section{Conclusions\label{sec:discussion}}
In this work we addressed the collective sedimentation of squirmers in a gravitational field concentrating on the relevant case, 
where the corresponding passive particles would completely sediment.
We showed that the sedimentation profile can be divided into three distinct regions;
close packed layers  near the bottom, a transition region, and a rather dilute active suspension at the top.
Here, the density depends exponentially on height, as for passive colloids, and the profile is very dynamic. 
The exponential dependence is non-trivial given the strong hydrodynamic interactions between the squirmers
due to their flow fields.
We also identified a strongly height-dependent mean orientation of the swimmers.
While the mean orientation is strongly varying across the close packed layers, in particular for small $|\beta|$,
it varies far less in the exponential region.
From the latter we extracted sedimentation lengths and showed that these not only grow with the ratio of active to sedimentation 
velocity, but also depend on the squirmer type.
We argued that neutral squirmers or weak pushers and pullers are more persistent when swimming upwards and thereby 
show larger sedimentation lengths.

Furthermore, sedimenting squirmers create strong convective currents due to their hydrodynamic interactions.
The spatial extension and the strength of convection are again determined by the rescaled swimming speed and 
by the squirmer type.
Neutral squirmers as well as weak pushers and pullers show the strongest convectional flow. 
In particular, for swimming speeds larger than the sedimentation velocity pronounced convection cells occur, 
which extend over the whole simulation box. 
Finally, as another signature of the highly dynamic sedimentation profile in the exponential region, 
we identified strong temporal fluctuations of the convective currents.

What we could not resolve in our current simulations due to limited computational resources is the question of what determines
the lateral extent of the convection cells. Is there an intrinsic length scale, which sets it? 
For this we would need to increase the simulation cell in the lateral directions. 
In future work, we plan to include bottom heaviness of the squirmers in our simulations and study in detail the inversion of the 
sedimentation profile, which was discussed in Ref.~\cite{Wolff:2013dm} for very dilute swimmer suspensions.
Preliminary results in denser systems show that instabilities occur due to hydrodynamic interactions~\cite{Hennes:V3vttE9l}.
In a harmonic trapping potential a similar instability leads to the formation of fluid pumps by breaking the rotational symmetry of the 
trap~\cite{Nash:2010ir, Hennes:2014hf}. 
In the present case we expect convectional patterns to occur.
Thereby, we will connect to the phenomenon of bioconvection~\cite{Wager:fp, Pedley:1992wa, Williams:2011cj, Williams:2011tt}.

\begin{acknowledgments}
We would like to thank C.~Cottin-Bizonne and A.~Z\"ottl for stimulating discussions.
This project was funded by Deutsche Forschungsgemeinschaft through the research training group GRK 1558
and priority program SPP 1726 (grant number STA352/11). 
Simulations were conducted at the ``Norddeutscher Verbund f\"ur Hoch- und H\"ochstleistungsrechnen'' 
(HLRN), project number bep00050.
\end{acknowledgments}

\bibliographystyle{apsrev4-1}
\bibliography{Collective_Sedimentation_of_Squirmers_under_Gravity}

\end{document}